\documentclass{article}

\usepackage{graphicx}

\usepackage{lipsum}
\usepackage{amsfonts}
\usepackage{epstopdf}
\usepackage{algorithm}
\usepackage{algorithmic}
\usepackage{subfigure}
\usepackage{booktabs}
\usepackage{color}
\usepackage{amsmath}
\ifpdf
  \DeclareGraphicsExtensions{.eps,.pdf,.png,.jpg}
\else
  \DeclareGraphicsExtensions{.eps}
\fi

\title{Domain-Decomposed Lagrangian Data Assimilation for Drifting Sea-Ice Floe Dynamics}

\author{Danyang Li\thanks{School of Computing, Australian National University, Canberra, ACT 2601, Australia 
  Email: Danyang.Li@anu.edu.au.}
\and  John Taylor\thanks{School of Computing, Australian National University, Canberra, ACT 2601, Australia 
  Email: John.Taylor@anu.edu.au}
\and  Quanling Deng\thanks{Yau Mathematical Sciences Center, Tsinghua University, Beijing, 100084, China Email: qldeng@tsinghua.edu.cn; School of Computing, Australian National University, Canberra, ACT 2601, Australia Email: Quanling.Deng@anu.edu.au.}}
\date{}

\begin{document}
\maketitle

\begin{abstract}
Sea ice dynamics are crucial to the global climate system, yet traditional continuum (e.g., viscous–plastic) models often fail to represent the discrete floe interactions that dominate in the marginal ice zone. Lagrangian discrete element methods (DEMs) resolve floe-scale physics more realistically, but their high particle counts make ensemble data assimilation (DA) more expensive. We consider a highly-simplified floe model and propose a scalable, domain-decomposed DA framework that couples Lagrangian particle observations with an ensemble transform Kalman filter (ETKF) to recover the underlying ocean flow field in a multiscale setting. The Eulerian domain is first partitioned into subdomains. We then impose an ETKF in each subdomain to recover the local fine-scale ocean features.  A Gaussian-weighted blending step then reconstructs a globally consistent flow field across subdomain boundaries. Numerical experiments demonstrate consistently better skill scores that are characterised by normalised root mean square error (NRMSE) and pattern correlation coefficients (PCC), compared to the global and expensive DA baseline. Results suggest that the domain-decomposed DA method is an alternative, scalable approach for particle-based sea-ice floe dynamics and ocean flow recovery.
\end{abstract}



\section{Introduction}
Sea ice covers large areas of the polar oceans and forms a dynamic interface between the atmosphere and ocean. Its seasonal advance and retreat modulate planetary albedo and help regulate global temperature \cite{kelman2017arcticness,kelman2022antarcticness,thomas2008sea}, while also influencing deep-water formation and the meridional overturning circulation \cite{serreze2007perspectives}.

At regional scales ($>100$ km), sea-ice drift is primarily driven by wind stress and ocean forcing \cite{comiso2008accelerated,meier2007whither}. Continuum models with viscous-plastic and elastic-viscous-plastic rheologies can reproduce large-scale motion \cite{bouillon2013elastic,hibler1979dynamic,hunke2001viscous,lemieux2009numerical}, but they homogenise floe-scale mechanics that become important below $\sim10$ km, where floes interact through collisions, cracking, and ridging \cite{cavalieri2012arctic,feltham2008sea,rampal2009positive}. With improving satellite products (e.g., SAR and ICESat) resolving features at $\sim$km scale \cite{comiso2008accelerated}, particle-resolving approaches are increasingly feasible. 

Lagrangian Discrete Element Methods (DEMs) evolve individual floes via Newton’s equations and contact mechanics \cite{damsgaard2018application,herman2016discrete,hopkins2004discrete}, enabling mechanistic representation of collision-driven dissipation, fracture, and floe-size redistribution \cite{damsgaard2018application,junior2021fully}. However, their cost grows rapidly with floe count, limiting applications to relatively small domains and idealised settings \cite{herman2016discrete}.

Data assimilation (DA) improves state estimation by combining observations with model dynamics 
\cite{law2015data}.  
In floe modelling, DA can adjust initial states and/or forcings to match observed Lagrangian trajectories \cite{chen2022superfloe,chen2021lagrangian,deng2025lemda}. Ensemble methods handle nonlinear effects by propagating multiple realizations of the system state, thereby capturing uncertainty and nonlinear evolution \cite{evensen2003ensemble}; the ensemble transform Kalman filter (ETKF) \cite{bishop2001adaptive} further reduces analysis cost, it is also commonly used with localisation, selective observation, and other practical extensions \cite{neal2009evaluating}. Nevertheless, scaling ETKF to Lagrangian DEM systems remains challenging due to high-dimensional states and dense observations.

Domain decomposition (DD) scales large numerical problems by splitting the domain into subdomains and solving coupled local problems
\cite{dolean2015introduction,quarteroni1999domain,toselli2004domain}. DD ideas have also been explored in ensemble DA to enable parallel updates, e.g., via subdomain-wise estimation of sparse inverse background-error structure \cite{nino2019parallel}. Building on these developments, we propose a DD strategy for assimilating sparse Lagrangian observations in sea-ice dynamical systems. We partition the Eulerian domain into non-overlapping subdomains and run ETKF within each, assimilating only base on in-subdomain floe observations.
Unlike fixed-radii localisation \cite{hunt2007efficient}, which may truncate long-range correlations in particle systems, we blend the updated subdomain fields using a Gaussian-weighted average to obtain a coherent global velocity field and suppress boundary artefacts.

The rest of the paper is organised as follows. Section~\ref{sec:main} formalises the problem and modelling assumptions. Section~\ref{sec:alg} presents the subdomain ETKF, floe selection, and Gaussian-weighted fusion. Section~\ref{sec:exp} reports NRMSE, PCC, and wall-clock runtime. Section~\ref{sec:conclusions} concludes and discusses limitations and future work.

\section{Particle model for free-drifting floe dynamics}
\label{sec:main}
We adopt an idealised Lagrangian DEM to describe sea-ice floe dynamics \cite{chen2022superfloe,chen2021lagrangian}.
Following standard practice \cite{herman2016discrete,damsgaard2018application}, each floe is represented as a cylinder of radius $r$ and thickness $h$. The floe mass is given by $m = \rho \pi r^{2} h$, where $\rho$ is the ice density. 
Consider a system of $L$ floes indexed by $l = 1,2,\ldots,L$. The position and velocity of floe $l$ are denoted by $\bold{x}_l=(x_l,y_l)$ and $\bold{v}_l=(u_l,v_l)$, and its mass, radius, and angular velocity are denoted by $m_l$, $r_l$, and $\omega_l$, respectively. 
We focus on the introduction of the domain-decomposed data assimilation technique in a multiscale setting and 
for the purpose of simplicity,  
we consider the dynamics of non-rotational free-drifting (thus non-collisional) sea ice floes. 
The modelling equations of free-drifting motion follow Newton’s second law and drag-forcing law:
\begin{subequations} \label{eq:dem}
\begin{align}
    &\frac{d\bold{x}_{l}}{d t} = \bold{v}_{l}, \label{eq:demx} \\
    &m_{l}\frac{d\bold{v}_{l}}{d t} = \alpha_{l}(\bold{u}^{o}-\bold{v}_{l})|\bold{u}^{o}-\bold{v}_{l}|, \label{eq:demv}  \\
    &\frac{d\bold{\hat{u}}^o}{d t} = (\bold{L^u\hat{u}}^o+\bold{F^u}) + \bold{\Sigma}^\bold{u}\bold{W}^{\bold{u}}(t), \quad \bold{u}^o =\bold{G(x)}\bold{\hat{u}}^o, \label{eq:democn} 
\end{align}
\end{subequations}
where $\boldsymbol{f}_{l} := \alpha_{l}(\bold{u}^{o}-\bold{v}_{l})|\bold{u}^{o}-\bold{v}_{l}|$ is the ocean drag force following the quadratic law, $\bold{u}^o$ is the ocean field, and $\bold{\hat{u}}^o$ is corresponding Fourier modes of the ocean.
Herein, we consider a one-way coupling in which a prescribed ocean current drives floe dynamics without feedback. This setup isolates the impact of the assimilated ocean state on floe motion and provides a clear baseline for validating the domain-decomposed technique.

The ocean flow field is represented in Fourier space: each mode $\hat{u}_{\bold{k},\zeta}$ is modelled as a linear stochastic system, where $\bold{k}$ is the 2D wavenumber and $\zeta$ indexes mode types associated with the same wavenumber \cite{berner2017stochastic,branicki2018accuracy,chen2022superfloe,farrell1993stochastic}. While the true dynamics involve nonlinear interactions, this linear stochastic approximation is widely used in DA to capture the dominant oscillatory and dissipative behaviour \cite{chen2022superfloe,chen2021lagrangian}, with additive noise parameterising unresolved processes and maintaining ensemble spread for uncertainty quantification.
With this in mind, a single Fourier mode is modelled as (see, for example, \cite{majda2006nonlinear})
\begin{equation}
    \frac{d\hat{u}_{\bold{k},\zeta}}{dt} = ((-d_{\bold{k},\zeta}+i\phi_{\bold{k},\zeta})\hat{u}_{\bold{k},\zeta}+f_{\bold{k},\zeta}) + \sigma_{\bold{k},\zeta}d{W}_{\bold{k},\zeta}.
\end{equation}\label{eq:motion4}
where $d_{\bold{k},\zeta}$ is the damping coefficient, $\phi_{\bold{k},\zeta}$ is the phase speed, $f_{\bold{k},\zeta}$ is the external forcing, $\sigma_{\bold{k},\zeta}$ is the strength of the stochastic forcing, and ${W}_{\bold{k},\zeta}$ is a complex-valued white noise.
Collecting all modes into the spectral state vector $\bold{\hat{u}}^{o}$, the compact representation of the ocean dynamics leads to \eqref{eq:democn}
where $\bold{L}^{u}$ is the linear operator representing damping and oscillation, $\bold{F}^{u}$ denotes external forcing, and $\bold{\Sigma}^{u} d\bold{W}^{u}$ is the multivariate stochastic forcing.
Finally, the ocean velocity field in physical space is reconstructed via the inverse Fourier transform:
\begin{equation}\label{eq:motion5}
    \bold{u}^o =\bold{G(x)}\bold{\hat{u}}^o,
\end{equation}
with $\boldsymbol{G}(\boldsymbol{x})$ denoting the inverse Fourier transform operator evaluated on the physical grid.

Despite this idealised setting, there are challenges in simulations, especially when moving toward floes with more features in complex ocean-floe dynamics. First, the computational cost increases rapidly with the number of floes $L$ since advancing the Lagrangian system requires evaluating the ocean velocity at each floe location. Second, the ocean flow field is not directly observed at the spatial and temporal resolutions required to drive floe-scale dynamics; in addition, available observation are often sparse and indirect.

\section{Domain decomposed ocean dynamics data assimilation}
\label{sec:alg}

\begin{algorithm}[!t]
    \begin{algorithmic}[1]
        \REQUIRE Total number of floes $L$; number of subdomains $S$; observations per subdomain $L^{\mathrm{obs}}$ (with $L \gg S \times L^{\mathrm{obs}}$).
        \ENSURE Updated local Fourier coefficients $\hat{\mathbf{u}}_{s}^{o}$ for each subdomain $A_s$.
        \STATE Initialise positions and velocities of $L$ floes; Initialise Fourier coefficients for the ocean model
        \STATE Partition the domain into $S$ non-overlapping subdomains $\{A_s\}_{s=1}^S$.
        \FOR {each subdomain $A_s$(parallelisable)}
            \STATE Compute the geometric centre $\mathbf{C}_s$ of $A_s$.
            \STATE Select $L^{\mathrm{obs}}$ floes within $A_s$ as the local observation set, giving priority to those with larger radii and closer to $\mathbf{C}_s$ (see Fig.~\ref{fig:selection}).
            \WHILE {assimilation time $t < T$} 
                \STATE Assimilate the observed floe positions using ETKF to update the local state vector (floe velocities and local Fourier coefficients $\hat{\mathbf{u}}_{s}^{o}$)
                \STATE $t \leftarrow t + \Delta t^{\mathrm{obs}}$
            \ENDWHILE
            \STATE Map $\hat{\mathbf{u}}_{s}^{o}$ to physical space to compute the local ocean velocity field $\mathbf{u}_s^o(\mathbf{x})$.
            \STATE Apply a Gaussian weighting function $W_s(\mathbf{x})$ centered on $A_s$ to $\mathbf{u}_s^o(\mathbf{x})$, storing the result for post-analysis fusion.
        \ENDFOR
        \caption{Domain-Decomposed Lagrangian Data Assimilation}
        \label{alg:overall}
    \end{algorithmic}
\end{algorithm}

\begin{figure}[!t]
    \centering
    \includegraphics[width=\linewidth]{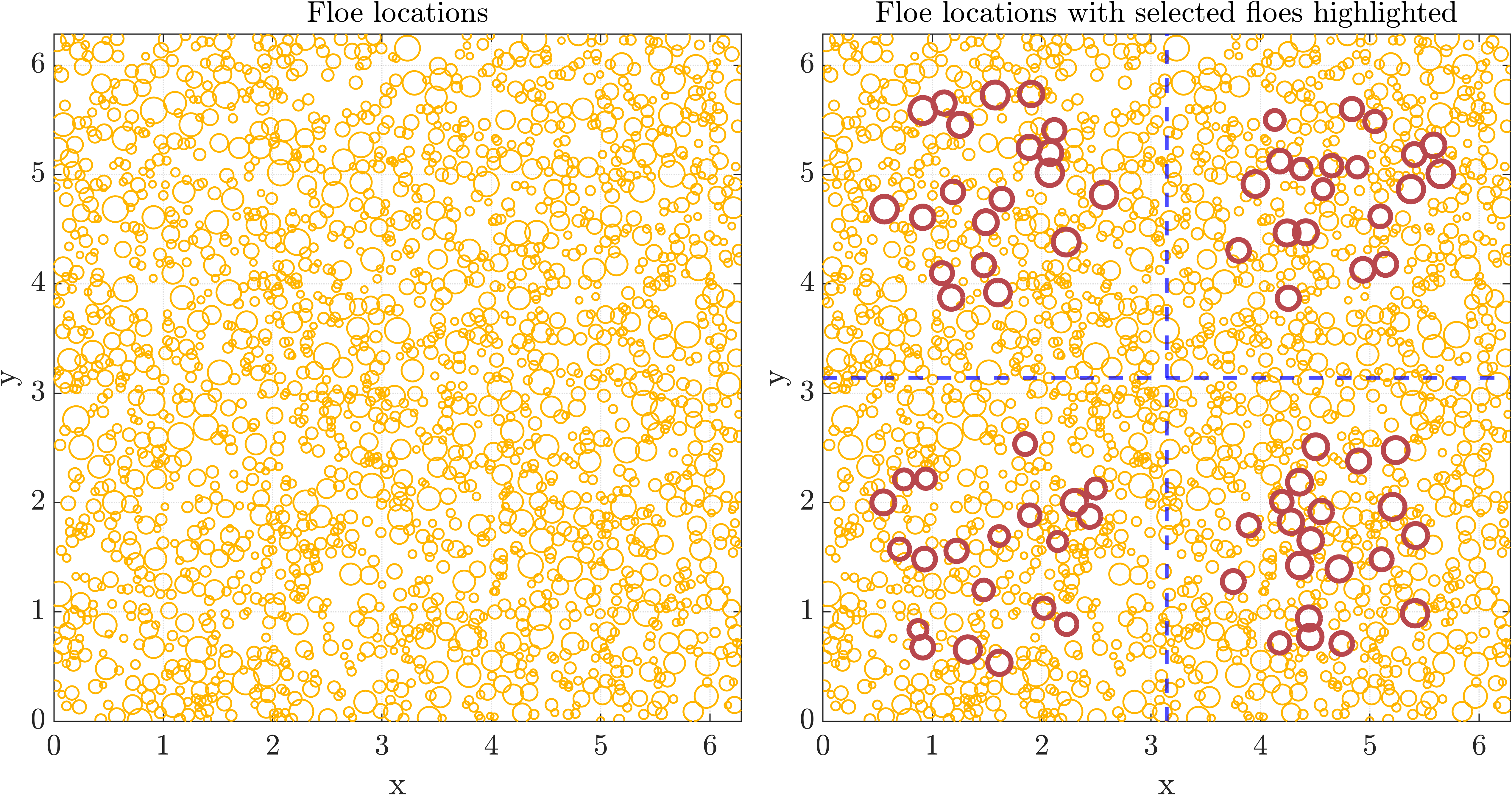}
    \caption{Example of local observation selection under domain decomposition.
    Red dots: the $L^{\mathrm{obs}}=20$ selected floes in each subdomain (chosen as the nearest and largest floes relative to the subdomain centre $C_s$); 
    yellow dots: all floes; 
    blue dashed lines: subdomain boundaries. 
    The domain is $[0,2\pi]\times[0,2\pi]$ with periodic boundary conditions.}
    \label{fig:selection}
\end{figure}

To assimilate dense, localised Lagrangian floe observations while recovering a coherent global ocean flow field, we propose a subdomain-based DA strategy (Algorithm~\ref{alg:overall}). The key idea is to (i) perform ensemble DA independently in each spatial subdomain using only local observations, and then (ii) reconstruct the global ocean field through smooth Gaussian blending of the recovered fields in subdomains.

We partition the original Eulerian domain $D$ into $S$ non-overlapping subdomains $\{A_s\}_{s=1}^S$ and process them independently (Algorithm~\ref{alg:overall}). In each $A_s$, we form a reduced local observation set by selecting $L^{\mathrm{obs}}$ representative floes, prioritising those closest to the subdomain centre $C_s$ (and, when applicable, larger floes), which reduces the observation dimension while retaining information most relevant to the local flow (Fig.~\ref{fig:selection}).
Given the selected observations, each subdomain performs a ETKF analysis to jointly update the local floe states and the subdomain ocean parameters (Fourier coefficients $\hat{\mathbf{u}}^{o}_{s}$). The updated coefficients are then mapped to physical space to obtain a local velocity field $\mathbf{u}^o_s(\mathbf{x})$ for subsequent Gaussian weighting and cross-subdomain fusion.

To suppress discontinuities across subdomain interfaces, we blend these subdomain fields using a smooth Gaussian weight matrix $W_s$ centred at each subdomain. 
Assume the domain is $D=[0,2\pi]\times[0,2\pi]$ with periodic boundary conditions for simulations. 
We decompose this domain uniformly into a grid of size $N_x \times N_y$. 
The subdomain sizes are $h_x = 2\pi/N_x$ for $x$ and $h_y=2\pi/N_y$ for $y$, respectively. 
The corresponding subdomain grid nodes (intersection of subdomain boundaries) are located at $(x_i, y_j), i = 1,2,\cdots, N_x, j= 1,2,\cdots, N_y.$
Specifically, we define
\begin{equation}
W_s(i, j) = \exp\left(-\frac{(x_i - C_s^x)^2 + (y_j - C_s^y)^2}{2(\sigma^{\mathrm{o}})^2}\right),
\end{equation}
where $(C_s^x, C_s^y)$ is the centre of subdomain $s$ and $\sigma^{o}$ 
controls the spatial decay. 
We then impose the normalised weights to ensure the partition of unity:
\begin{equation}
W_s^{\text{norm}}(i, j) = \frac{W_s(i, j)}{\sum_{s=1}^{S} W_s(i, j)}.
\end{equation}
The global ocean velocity field is then reconstructed by Gaussian-weighted fusion:
\begin{equation}
\mathbf{u}^{o}(x_i,y_j)
= \sum_{s=1}^{S} W_s^{\text{norm}}(i,j)\,\mathbf{u}^o_s(x_i,y_j)
\end{equation}
where $\mathbf{u}^{o}(x_i,y_j)$ is the nodal value and for an arbitrary point $(x,y) \in D$, $\mathbf{u}^{o}(x,y)$ is constructed using the bilinear finite element basis as an interpolation. 
This decomposition reduces the computational burden relative to full-domain assimilation in high-dimensional settings with many floes and Fourier modes, and it is naturally parallelisable because subdomain assimilations are independent. Within each subdomain, ETKF provides an ensemble-space analysis transform, enabling more accurate local updates before global reconstruction.

\section{Experimental Results}
\label{sec:exp}
\begin{table}[!t]
\centering
\renewcommand{\arraystretch}{1.15}
\begin{tabular}{@{}lll@{}}
\toprule
\textbf{Quantity} & \textbf{Non-dimensional} & \textbf{Physical scales} \\
\midrule
Computational domain
& $D = [0, 2\pi] \times [0, 2\pi]$
& $ \approx 50 \times 50$ km$^{2}$ \\
Floe radiis
& $r \in [0.004, 0.016]$
& $ \approx [32, 127] \text{ m}$\\
Max. ocean current speed
& $u^{\text{ocean}}_{\max} \approx 2$
& $\approx 0.18 \text{ m/s}$ \\
Max. ice speed
& $u^{\text{ice}}_{\max} \approx 5$
& $\approx 0.46 \text{ m/s}$ \\
Time step (model)
& $\Delta t = 1 \times 10^{-3}$
& $\approx 86.4 \text{ s}$ \\
Observation interval
& $\Delta t^{\text{obs}} = 1 \times 10^{-2}$
& $\approx 14.4 \text{ min}$ (every 10 steps) \\
\bottomrule
\end{tabular}
\vspace{0.3cm}
\caption{Non-dimensionalisation, units, and run configuration.}
\label{tab:nondim_units}
\end{table}

\begin{table}[!t]
\centering
\renewcommand{\arraystretch}{1.15}
\begin{tabular}{@{}lllll@{}}
\toprule
\textbf{Method} & \textbf{Grid size} & \textbf{Per-subdomain $L^{\mathrm{obs}}$} & \textbf{Total $L^{\mathrm{obs}}$} \\
\midrule
Full-domain & $1\times1$ & -- & 20,\;50,\;100,\;200 \\
\addlinespace
Decomposed & $2\times2$ & 10,\;20,\;50,\;100 & 40,\;80,\;200,\;400 \\
           & $4\times4$ & 5,\;10,\;20,\;50 & 80,\;160,\;320,\;800  \\
\bottomrule
\end{tabular}
\vspace{0.3cm}
\caption{Observation configurations used in the experiments. For decomposed runs, the total number of observed floes equals $S \times L^{\mathrm{obs}}$.}
\label{tab:obs_configs}
\end{table}

\begin{table}[!t]
\centering
\small
\begin{tabular}{lrrrr}
\toprule
Grid size & $L^{\mathrm{obs}}$ (floe/subdom.) & NRMSE $\downarrow$ & PCC $\uparrow$ & Runtime ($10^3$ s) $\downarrow$ \\
\midrule
$1\times1$ & 20  & 0.90 & 0.48 & 4.6 \\
$1\times1$ & 50  & 0.91 & 0.50 & 5.9 \\
$1\times1$ & 100 & 0.70 & 0.74 & 7.7 \\
$1\times1$ & 200 & \textbf{0.68} & \textbf{0.77} & 12.0 \\
\addlinespace
$2\times2$ & 10  & 0.83 & 0.57 & 4.6 \\
$2\times2$ & 20  & 0.75 & 0.68 & 4.3 \\
$2\times2$ & 50  & 0.65 & 0.76 & 7.5 \\
$2\times2$ & 100  & \textbf{0.55} & \textbf{0.83} & 8.0 \\
\addlinespace
$4\times4$ & 5  & 0.89 & 0.62 & 4.3 \\
$4\times4$ & 10  & 0.85 & 0.67 & 4.3 \\
$4\times4$ & 20  & 0.82 & 0.68 & 4.2 \\
$4\times4$ & 50  & \textbf{0.74} & \textbf{0.74} & 5.4 \\
\bottomrule
\end{tabular}
\vspace{0.3cm}
\caption{Final-time accuracy and runtime by grid size and observation density.}
\label{tab:tradeoff(k9)}
\end{table}

We consider the dynamics of $40{,}000$ free-drifting floes of unit thickness, with radii sampled from a power-law distribution $N(r)\propto r^{-\alpha}$ with $\alpha=1.3$. 
The ocean flow is represented by truncated GB modes with $k_{\max}=9$; each mode evolves under linear damping and stochastic forcing with coefficients $d=0.5$ and $\sigma=0.05$. 
Non-dimensional variables are mapped to physical units as in Table~\ref{tab:nondim_units}, and all experiments are run for $T=20$ days. 
Data assimilation is performed using an ETKF with 1000 ensemble members. 
Observations are noisy floe positions $(x,y)$ with Gaussian errors (standard deviation $0.01$ in non-dimensional units). 
We set $\sigma^{o}=2.6$ empirically based on numerical experiments. We remark that alternative reconstruction and smoothing procedures for the ocean field reconstruction could be considered for potentially improved performance but this is not the main focus of this work.
We compare a full-domain baseline against the proposed domain-decomposed scheme with grids $S\in\{2\times2,\,4\times4\}$, observation budgets are summarised in Table~\ref{tab:obs_configs}.
After ETKF updates, subdomain velocity fields are reconstructed in physical space using Gaussian weights over overlapping regions to recover a globally coherent field.

Table~\ref{tab:tradeoff(k9)} summarises results for the $1\times1$ full-domain baseline and decomposed $2\times2$ and $4\times4$ grids, across a range of observation counts $L^{\mathrm{obs}}$ (per subdomain).
Overall, increasing $L^{\mathrm{obs}}$ improves accuracy, with the largest gains from sparse to moderate observation budgets and diminishing returns thereafter. For the baseline, NRMSE decreases from 0.90 ($L^{\mathrm{obs}}=20$) to 0.70 ($100$) and 0.68 ($200$), while PCC increases from 0.48 to 0.74 and 0.77. A similar pattern holds for $2\times2$: raising $L^{\mathrm{obs}}$ from 10 to 50 per subdomain improves NRMSE/PCC from 0.83/0.57 to 0.65/0.76, with smaller additional gains at $L^{\mathrm{obs}}=100$ (0.55/0.83). The trend is weaker for $4\times4$, consistent with fewer observations per local region.
For comparable observation budgets, $2\times2$ consistently outperforms $4\times4$ at the final time, suggesting that finer partitioning becomes information-constrained under current tuning: smaller subdomains capture cross-boundary correlations less effectively and local updates recover large-scale coherence less well. Relative to $1\times1$, the $2\times2$ setting can match or exceed accuracy with fewer observations. 

If the grid is further refined (larger $S$), each subdomain contains fewer floes; with a fixed $L^{\mathrm{obs}}$, the selected representative floes are therefore more likely to lie near subdomain interfaces. In that case, their trajectories may be driven by flow structures that extend across neighbouring subdomains. While our Gaussian-weighted blending mitigates some cross-boundary inconsistencies, it cannot fully recover missing cross-subdomain information, which can limit further accuracy gains as partitioning becomes finer. 
Moreover, increasing the $L^{\mathrm{obs}}$ raises computational cost substantially, since the model must propagate a larger Lagrangian state by computing drag forcing for every floe at each time step.

All wall-clock runtime ranges from $\sim4.2\times10^3$ to $1.2\times10^4$ s and increases with $L^{\mathrm{obs}}$. Decomposition can achieve similar or higher accuracy at lower cost; for example, with 200 floes observed, $2\times2$ with $L^{\mathrm{obs}}=50$ attains NRMSE/PCC $=0.65/0.76$ in $7.5\times10^3$ s, compared with $0.68/0.77$ in $1.2\times10^4$ s for $1\times1$.

\begin{figure}[!t]
    \centering
    \includegraphics[width=\linewidth]{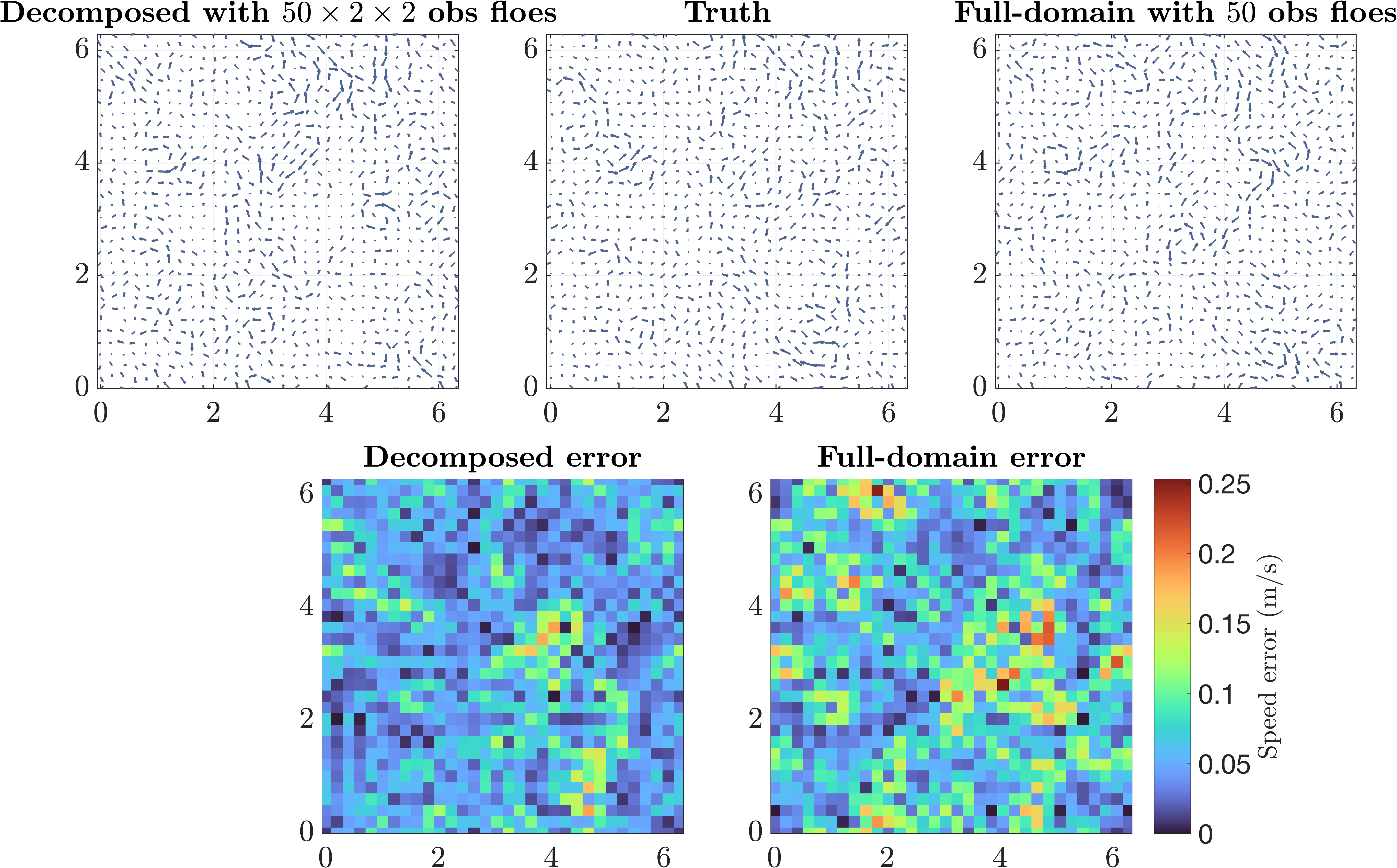}
    \caption{Ocean velocity-field reconstructions and error distributions at $T=20$. Top row: velocity fields from the decomposed DA (left), ground truth (middle), and full-domain DA (right). Bottom row: error magnitude maps comparing the decomposed (left) and full-domain (right) reconstructions against the ground truth.}
    \label{fig:figk9}
\end{figure}

To complement these statistics, we next visualise a representative snapshot of the reconstructed flow field. Figure~\ref{fig:figk9} compares the velocity-field reconstruction at day 20 for decomposed and full-domain DA. The decomposed method aligns closer with the ground truth, particularly in preserving localised circulation features. The corresponding error-magnitude maps confirm this improvement, with fewer high-error regions and reduced error magnitudes in the decomposed case.

\section{Conclusions}
\label{sec:conclusions}

In this paper, we developed a domain-decomposed DA framework for multiscale ocean field recovery in sea-ice modelling that consistently outperforms a full-domain baseline across key metrics.
In our simplified system without contact force, for NRMSE, the method achieves 0.75 ($L^{\mathrm{obs}}=20$) and 0.65 ($L^{\mathrm{obs}}=50$), exceeding baseline results of 0.9 and 0.91, demonstrating a 16–28\% improvement in accuracy; for PCC, the decomposed method achieves 0.68 and 0.76 while baseline results are 0.48 and 0.50, demonstrating a $\sim 50\%$ improvement.
These improvements arise from three synergistic mechanisms: (1)  each subdomain provides a finer‑scale view of the floe field, preserving details that are often lost in large-scale treatment in a full‑domain; (2) the subdomains are independent, their updates can be computed in parallel; and (3) Gaussian‑weighted blending keeps large‑scale flow structures smooth while damping noise at subdomain edges. 

As further work, we will extend the framework to fully coupled DEM configurations with floe rotations and floe--floe contact mechanics. In such settings, we expect the domain-decomposed analysis to be more advantageous because contact-induced stresses and deformation are strongly localised in space, performing inference within subdomains may better preserve these local interactions while maintaining global consistency through blending.

Overall, the proposed framework supports observationally economical and regionally adaptive DA for sea-ice dynamics and related high-dimensional multiscale geophysical systems, providing a scalable and alternative for accurate Lagrangian state estimation under limited observations.

\bibliographystyle{unsrt}
\bibliography{ref}

\end{document}